\documentclass[manuscript]{aastex}

\begin{document}

\title{Eclipses by a Circumstellar Dust Feature\\
  in the Pre-Main Sequence Star KH15D}

\author{Catrina M. Hamilton\altaffilmark{1,2} and William Herbst}
\affil{Astronomy Department, Wesleyan University, Middletown, CT
06459}
\email{catrina@astro.wesleyan.edu,bill@astro.wesleyan.edu}

\author{Candice Shih}
\affil{Department of Astronomy, Haverford College, Haverford, PA
19041}
\email{cshih@haverford.edu}

\and

\author{Anthony J. Ferro}
\affil{University of Arizona, Steward Observatory/NICMOS, Tucson, AZ
85721}
\email{tferro@as.arizona.edu}

\altaffiltext{1}{Physics Department, Wesleyan University, Middletown, CT 06459}
\altaffiltext{2}{Department of Physics, Astronomy, and Geophysics,
Connecticut College, New London, CT 06320; cmham@conncoll.edu}

\begin{abstract}

Photometry and spectroscopy of the unique pre-main sequence eclipsing
object KH15D in the young cluster NGC 2264 are presented. The orbital
period is 48.34 days and both the length ($\sim$16 d) and depth ($\sim$3 mag)
of the eclipse have increased with time. A brightening near the time
of central eclipse is confirmed in the recent data but at a much
smaller amplitude than was originally seen. Spectra taken
when the star is bright show that the primary is a weak T
Tauri star of spectral type K7. During eclipse there is no detectable change in
spectral type or reddening, indicating that the obscuration is
caused by rather large dust grains and/or macroscopic objects.
Evidently the star is eclipsed by an extended feature in its
circumstellar disk orbiting with a semi-major axis of $\sim$0.2 AU. 
Continued photometric monitoring
should allow us to probe the disk structure with a spatial resolution of
$\sim$3 x 10$^{6}$ km or better.
\end{abstract}
 
 
\section{Introduction}

T Tauri stars are recognized as pre-main sequence (PMS) stars 
(e.g. Bertout 1989) and exhibit photometric variability.  Their 
light variations are believed to arise primarily from a changing pattern of hot 
accretion zones and cool spots on their surfaces, but 
variable extinction caused by circumstellar matter may be involved
in some instances (Herbst et al. 1994, Bertout 2000).  
The presence of circumstellar disks around PMS stars has been inferred 
since the late 1960s.  Shu, Adams, and Lizano (1987) proposed the penultimate 
evolutionary stage of low-mass star formation as a T Tauri star with a 
surrounding remnant nebular disk.  The disappearance of the nebular disk 
occurs as matter accretes onto the star, becomes incorporated into planets
or stellar companions, or is dispersed by an energetic outflow.  Circumstellar 
disks have now been directly imaged by the Hubble Space Telescope 
(O'Dell \& Wen 1994, O'Dell \& Wong 1996).  With recent detections of
Jupiter-sized planets orbiting main-sequence stars at distances of 0.2 - 3 AU 
(Marcy \& Butler 2000), the question remains: how do
disks surrounding young stellar objects evolve and, in particular, 
what is the time
scale for planet formation?  Direct observation of circumstellar disk 
structure could place strong constraints on the nature of physical processes 
occuring in protoplanetary disks but most PMS stars are too distant to allow
much resolution of their disks with current imaging capabilities.

During a photometric study of the young cluster NGC 2264, Kearns 
\& Herbst (1998; hereafter KH) discovered a unique and
potentially important object, which we call KH15D, following their
designation.  It is a periodic variable (P $\sim$ 48 days) that appears
to undergo a regular eclipse, but with an amplitude of 3 to 4 magnitudes
in the Cousins {\it I} band.  Amazingly, the duration of the eclipse
is about one-third
of the period, or 16 days.  The star has now been observed photometrically
through dozens of cycles and its period and phase are well known
and predictable, indicating that we are dealing with an orbital phenomenon.
The depth of the eclipse, its duration, and the evidence for photometric
variability during totality all indicate that the eclipsing object is not
a normal star.  Here we present and discuss evidence that KH15D is being
eclipsed by a feature in its circumstellar disk, and describe the constraints
our data place on the nature of that feature.

\section{Observations}

\subsection{Photometry}

Photometry of KH15D was obtained during six observing seasons from
1995 October through 2001 April with the 0.6 m Perkin telescope of 
the Van Vleck
Observatory on the campus of Wesleyan University.  A Cousins {\it I} filter was
used for all observations and exposure times were 5 minutes.  The
scale of the images is 0.6$\arcsec$ per pixel and typical seeing was about
2.5$\arcsec$.  Flat-fielding was accomplished with sky flats and the images
were also corrected for bias and dark noise.  Aperture photometry was performed
using standard IRAF\footnote{Image Reduction and Analysis Facility,
written and supported by the IRAF programming group at the National
Optical Astronomy Observatories (NOAO) in Tucson, Arizona.  NOAO is
operated by the Association of Universities for Research in Astronomy (AURA),
Inc. under cooperative agreement with the National Science Foundation.}
tasks and an aperture diameter of 4.8$\arcsec$.
Differential magnitudes were determined by reference to a set of comparison
stars on the same frame.  Typical errors are $\sim$0.01 mag when the 
star is in
its bright state and $\sim$0.1-0.2 mag when in eclipse.  Portions of frames
showing KH15D in and out of eclipse are shown in Fig.  1 to identify the
star and to illustrate its extraordinary range.

A periodogram analysis of the full data set indicates that the star's
period is 48.35 d, consistent with the value of 48-49 d reported by KH based on
two seasons of data.  This was modified slightly, to 48.34 d $\pm$ 0.02
(estimated error) by inspection of the folded light curves.  Phased light
curves at three epochs of observation are shown in Figure 2.  
It is evident
that the shape of the light curve has evolved substantially over the five
year interval represented by the data.  In particular, the central brightening 
in the first two epochs appears
to be significantly brighter than the out-of-eclipse levels at those
epochs.  In 95-96, the magnitude of this effect
is $\sim$0.5 mag and in 96-97 it is $\sim$0.1 mag.  By 99-00 however, 
the central reversal has diminished dramatically and is well below the
out-of-eclipse level for that year.  
There may also be a slight lengthening of the eclipse duration between 96-97 
and 99-00 of about 0.02 in phase ($\sim$1 day) but it is difficult to be 
certain given the gaps in the light curve.  
We are perhaps seeing the evolution of 
a feature driven by orbital dynamics, although the source of the dramatic
central brightness elevation in 95-96 remains mysterious.
An ephemeris for mid-eclipse based on these data
is $$JD(mid-eclipse) = 2451626.86 + 48.34E.$$
Out of eclipse the star appears to be of constant brightness to within 
$\pm$ 0.03 mag.

\subsection{Spectroscopy}

Moderate resolution spectra ($\Delta\lambda \sim$ 4 \AA) of KH15D
were obtained on the nights of 15, 16, and 17 December 2000 when it
was in its bright state (phase = 0.56)
and again on 14 January 2001 during an eclipse (phase = 0.15).
The data were collected using the B and C spectrograph plus
the Loral 800 x 1200 CCD detector on Steward Observatory's 2.3 meter
telescope through light, variable cirrus clouds.
These moderate resolution spectra were taken using the 600 lines per mm
grating blazed at
6681 \AA, covering the wavelength range $\sim$ 4860 - 7000 \AA, at a dispersion
of 1.83 \AA/pixel. The L42 filter was used to block light with 
wavelengths shorter
than 4200 \AA. The images were processed using standard IRAF tasks.
This included bias subtraction and flat field corrections using a
quartz lamp. All frames were found to have a slight gradient present along
the slit.  In order to remove this, an illumination
correction was performed.  Dome flats were used to determine the slit
illumination function.  This was accomplished with the task ILLUM in
IRAF.  One-dimensional
spectra were then extracted using the APALL
task, and wavelength calibration was established from exposures of a He-Ar
comparison lamp taken immediately following each program star exposure.  Lastly, all spectra were flux calibrated using the standard star G191b2b and the usual IRAF tasks (STANDARD, SENSFUNC, CALIBRATE). 

Figure 3 shows our reduced spectra for KH15D in and out of eclipse.  The
spectrum in the top panel is a 1 hr exposure of the star in its bright
state.  The spectrum in the bottom panel is an average of six one hour
exposures obtained during eclipse.  We multiplied this spectrum by 14.7 so 
that both spectra could appear on the same scale.  
Locations where a bad background and night sky subtraction could not
be avoided are marked.  These are labeled as BGS and NS respectively.
Atmospheric absorption bands are identified by ATM.
Direct comparison of 
the spectrum 
in the bright phase was made with standard star spectra obtained using the same 
telescope and instrument at a lower resolution for classification 
purposes.  Careful inspection of the 
spectra in both panels shows that they are nearly identical, and that they
both exhibit stellar features indicative of a K7 V spectral type.
In particular, the presence and strength of the TiO features, in addition 
to the MgH feature, is clear evidence of a late K star.  Other lines used 
for classification purposes were chosen from the list given by Cohen \&
Kuhi (1979) and are labeled in the plot. 
The presence of the $\lambda$6707 Li I feature and the lack of any 
substantial H$\alpha$ emission indicate that this is a weak-lined T 
Tauri star (WTTS) and confirms the youthfulness of this system.  
During the bright phase, the Li feature has an equivalent
width of 0.469 \AA, consistent with the Li equivalent widths measured
for stars in NGC 2264 by Soderblom et al. (1999).  It should be noted that 
due to the low resolution of 
our spectra, the Li feature is blended with the Fe I line at 6707.441 \AA,
and therefore our measured value should be taken as an upper limit.   
A cross-correlation of the spectrum in eclipse with the spectrum out of 
eclipse was performed using the FXCOR task in IRAF in an effort to determine
a radial velocity.  The spectra were not sufficiently resolved, however, to
deduce a reliable relative radial velocity.

\subsection{Properties of the Star}

Table 1 lists the known and inferred properties of KH15D.  Photometry
presented has been obtained by two different groups. 
We used Park et al.'s (2000) distance to NGC 2264, and their 
values of V and B-V, to determine the absolute magnitude of KH15D.  A value 
of $M_{bol}$ was then obtained by applying the bolometric correction 
used by Hillenbrand (1997) for stars with 3.826 $\geq$ log $T_{eff}$ 
$\geq$ 3.571.
Adopting $M_{bol,\odot}$ = 4.74 (Bessel et al. 1998), a luminosity and radius 
for this star was calculated.  We then used evolutionary tracks and
isochrones from  D'Antona \& Mazzitelli (1994, 1998), and Baraffe et al. (1998) 
to obtain estimates for the mass and age of KH15D.
Out of eclipse, the star appears to be an ordinary WTTS member of
the young cluster NGC 2264 with a mass of $\sim$0.5-1.0 $M_{\odot}$ and an 
age of $\sim$2-10 My.
The large range in these values reflects the scatter in the models used.
In summary, KH15D appears to be a normal PMS star. 
We suppose that what is unusual about 
this star is that it has a fortuitous alignment of its circumstellar disk 
with our line of sight such that an orbiting feature in the disk periodically
occults the star.

\section{Discussion}

KH15D exhibits unique photometric behavior, to the authors' knowledge.  The
strict periodicity, large amplitude and shape of the light curve make it
certain, in our view, that we are witnessing the eclipse of a star by an
orbiting extended body (or ``feature").   
Since we were unable to determine accurate radial
velocities of the star from our spectra, we do not yet know for certain
whether the K7 star is the more massive object, but that
seems likely for the following reason.  If the unseen object were more
massive, and the visible star were being occulted by a disk surrounding it,
that disk would need to extend nearly the entire distance between the
components to account for the observed length of the eclipse ($\sim$1/3 of
the orbital period), assuming a circular orbit.  This is certainly
not a dynamically stable arrangement.  Circumstellar disks
in a binary system are limited to radii of $\sim$ 0.2-0.5a, where a is the 
closest approach of the components (Mathieu et al. 2000).  Assuming a 
circular orbit, this would limit the length of an eclipse in
such a system to $\leq$ 7 d, much shorter than observed.  Also, we note
that there is no evidence of a second source of light in the system, as
might be expected if the unseen feature were more massive than the K7 star.  Therefore, we expect that radial velocity data will show little or no
orbital motion for the K7 star, although this clearly needs to be checked.
Very high precision data might ultimately reveal the mass of an unseen
component, if it is comparable to that of a giant planet.

Assuming, therefore, that the occulting body is of relatively low mass and
orbiting the K7 star, and adopting the properties of the system given in
Table 1, we find that the semi-major axis of the orbit is 0.2 AU, or about
22 stellar radii and that the feature extends over about 1/3 of the
circumference of the orbit, assuming it is not highly eccentric.  Clearly
it must consist of a myriad of smaller
objects that are larger than interstellar dust grains (since no reddening
is associated with the obscuration) but still have substantial surface area
for their mass (to produce the required extinction).  The
amazing reversals seen in the 1995-1996 and 1996-1997 seasons, when the star 
exceeded its usual 
bright state, indicate that there are substantial density
fluctuations within this obscuring cloud and, sometimes, holes.  The
leading and trailing edges seem to be fairly sharply defined.  A knife edge
occulting the star would lead to an ingress and egress rate of about 0.27
mag/hr, which would last about 0.45 day.  These values are consistent with
what we actually observe and indicate, therefore that the edge of the
occulting feature is rather sharp.

In summary, KH15D appears to be a 2-10 Myr old, 0.5-1.0 M$_{\odot}$ star 
orbited and periodically occulted by an extended dust feature at a distance of 
about 0.2 AU.  The dust feature is obviously asymmentric, structured
and evolving on timescales of months to years.  It extends $\sim$ 120$\degr$
and could be a ``ring arc" similar to what is seen in Neptune or a portion
of a warped disk that extends vertically into the line of sight only at
certain longitudes.  Perhaps it has its origin in the interplay of the 
star's magnetic field with the disk, analogous to what has been proposed 
for AA Tau by Terquem \& Papaloizou (2000).  However, this feature is much
farther out from the star than in that case, applies to a weak T Tauri star,
in which the gaseous disk is likely to be already absent, and the star has 
no evidence of a particularly strong, well-aligned magnetic field (i.e., no
spot variations outside of eclipse).  Therefore, we consider it more 
probable that 
the source of the warp or asymmetry in the disk is gravitational
in nature.  It is, perhaps, of note that the onset of the feature is at $\pm$
60\degr{\ } relative to the center of the feature, namely at the Lagrangian
points of the orbit of a putative central body.  The vertical extent of the 
feature at its center must be comparable to or larger than the star, 
of course, since 
there is nearly complete blocking of the light at the most recent epoch. 
However, the fact that there is a central reversal of the light to its full
(or even somwhat exceeding its full) brightness in the earlier epochs suggests
that the scale height may not be much larger than the star.  It is probably
not warranted to speculate further about this object at present.  Clearly,
the star deserves special attention because of its potential to yield 
information on a spatial scale that will not be reached by imaging for many 
years to come.  We are organizing an observing campaign for the upcoming season
and would welcome participation by any interested parties.

\acknowledgements

We would like to thank the many students who contributed to the 
observing program over the years, especially Kate Eberwein, a Vassar
student, and Frank Muscara, a Wesleyan student.  One of us (CS) was
supported by a summer research fellowship from the Keck Northeast 
Astronomy Consortium, which is funded by a grant from the W.M. Keck 
foundation.  One of us (WH) acknowledges support from NASA through its
Origins of Solar Systems program.  One of us (CMH) was supported in 
part by a grant from the R.F. Johnson Faculty Development Fund through
Connecticut College.

\clearpage

\begin{deluxetable}{ll}
\tabletypesize{\footnotesize}
\tablecaption{Relevant Properties of KH15D. \label{tbl-1}}
\tablewidth{0pt}

\startdata
Parameter & Value\\
Position: &$\alpha$ = 06:41:10.18, $\delta$ = +09:28:35.5, 2000.0\\
Alternate Identifications: &150\tablenotemark{a}, 391\tablenotemark{b}\\
Photometry (Out of Eclipse): &V = 16.096\tablenotemark{a}, 16.00\tablenotemark{b}\\
\nodata &U-B = 1.081\tablenotemark{a}\\
\nodata &B-V = 1.308\tablenotemark{a}, 1.32\tablenotemark{b,c}\\
\nodata &V-I = 1.571\tablenotemark{a}\\
\nodata &V-R = 0.82\tablenotemark{b}\\
\nodata &R-I = 0.80\tablenotemark{b}\\
Eclipse Depth: &3 mag in {\it I}\\
Eclipse Ephemeris: &$JD(mid-eclipse)=2451626.86 + 48.34E$\\
Spectral Type: &K7 V\\
W$_{\lambda}$(Li): &0.469 \AA\\
Log $T_{eff}$: &3.61\tablenotemark{b}, 3.602\tablenotemark{d}\\
Reddening, Extinction: &0, 0\\
Bolometric Correction: &-0.97\tablenotemark{e}\\
Distance: &760 pc\tablenotemark{a}, 760 pc\tablenotemark{f}\\
Luminosity: &0.5 ($L_{\odot}$)\tablenotemark{b}, 0.4 ($L_{\odot}$)\\
Radius: &1.3 ($R_{\odot}$)\\
Mass: &0.6 ($M_{\odot}$)\tablenotemark{b}, 0.5-1.0 ($M_{\odot}$)\tablenotemark{g}\\
Age: & 2 Myr\tablenotemark{b}, 2-10 Myr\tablenotemark{h}, \\

\enddata

\tablenotetext{a}{Park et al. 2000}
\tablenotetext{b}{Flaccomio et al. 1999}
\tablenotetext{c}{The B-V value originally quoted in Flaccomio et al. 1999 was 0.32.  Since this is grossly different from what is expected for this star, we assumed that the value was supposed to be 1.32.}
\tablenotetext{d}{Value obtained from Cohen \& Kuhi 1979.}
\tablenotetext{e}{Hillenbrand 1997}
\tablenotetext{f}{Sung et al. 1997}
\tablenotetext{g}{Values obtained using evolutionary tracks based on PMS
evolution models from D'Antona \& Mazzitelli (1994, 1998) and Baraffe et al. (1998).}
\tablenotetext{h}{Values obtained using evolutionary tracks based on PMS
evolution models from D'Antona \& Mazzitelli (1994, 1998) and Baraffe et al. (1998).}

\end{deluxetable}

\clearpage

\begin{figure}
\plottwo{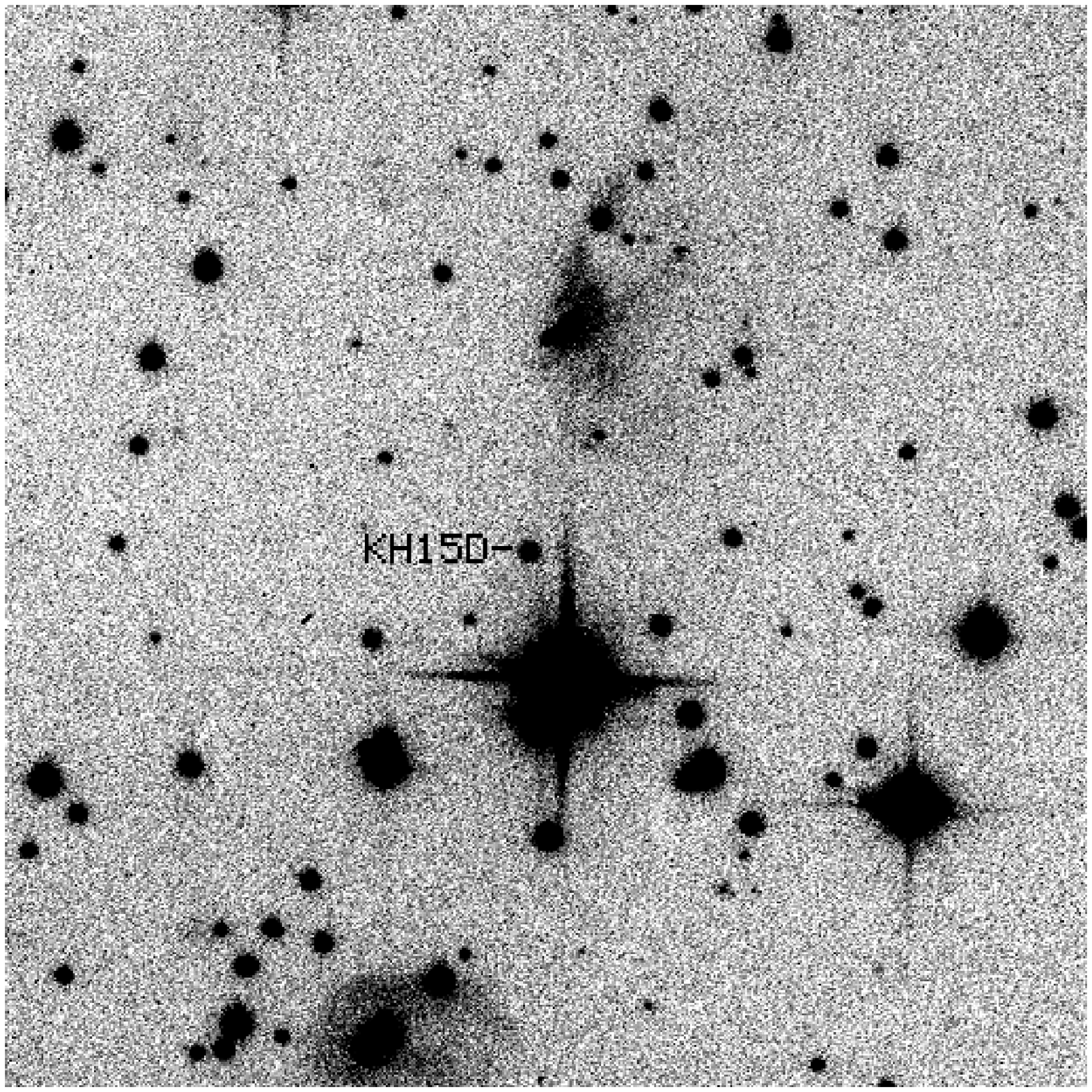}{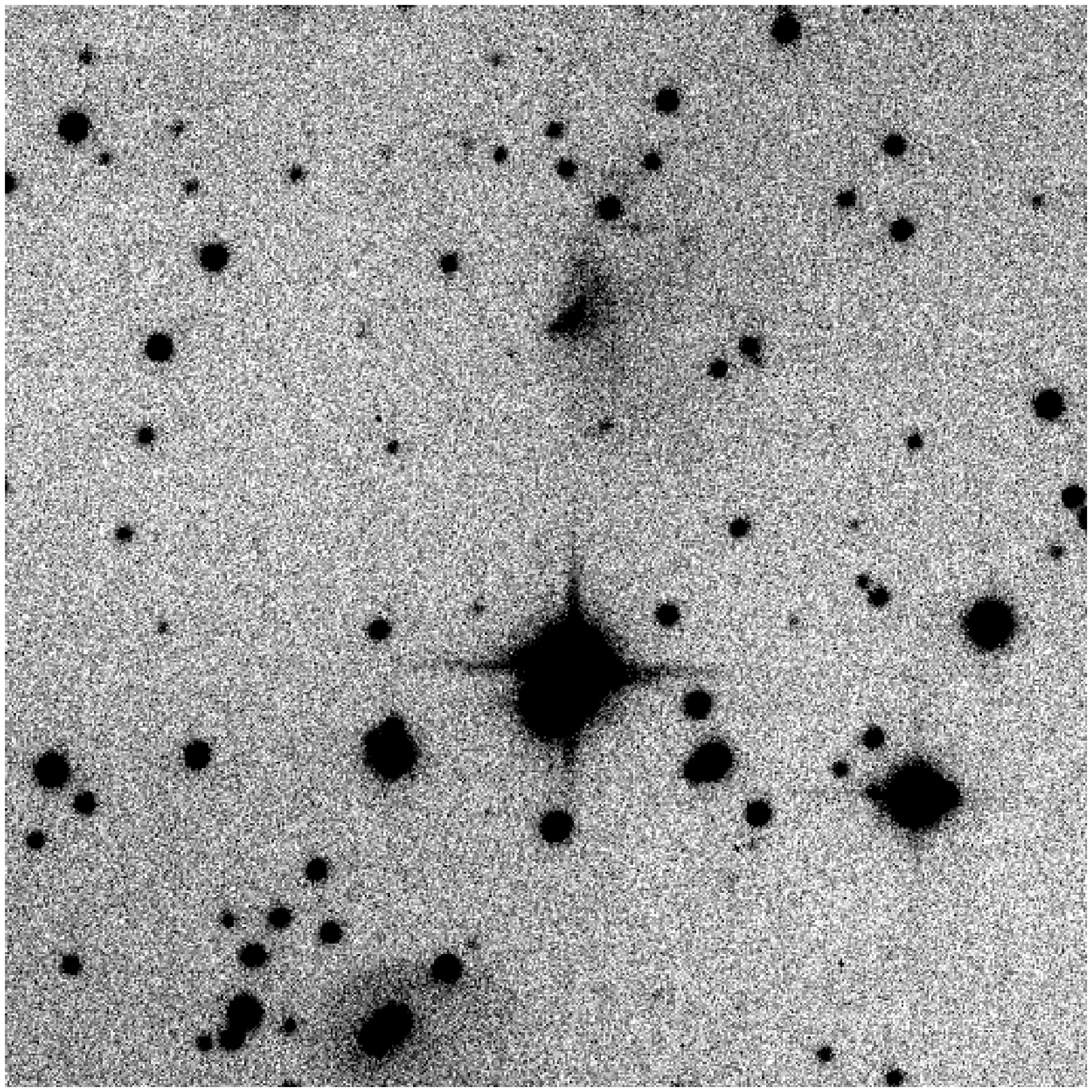}
\figcaption{Images of KH15D in and out of eclipse obtained through a
Cousins I filter with the 0.6 m telescope at Van Vleck Obs. North is
at the top and east to the left and the frames are $\sim$5$\arcmin$
on a side. \label{Fig. 1}}
\end{figure}

\begin{figure}
\plotone{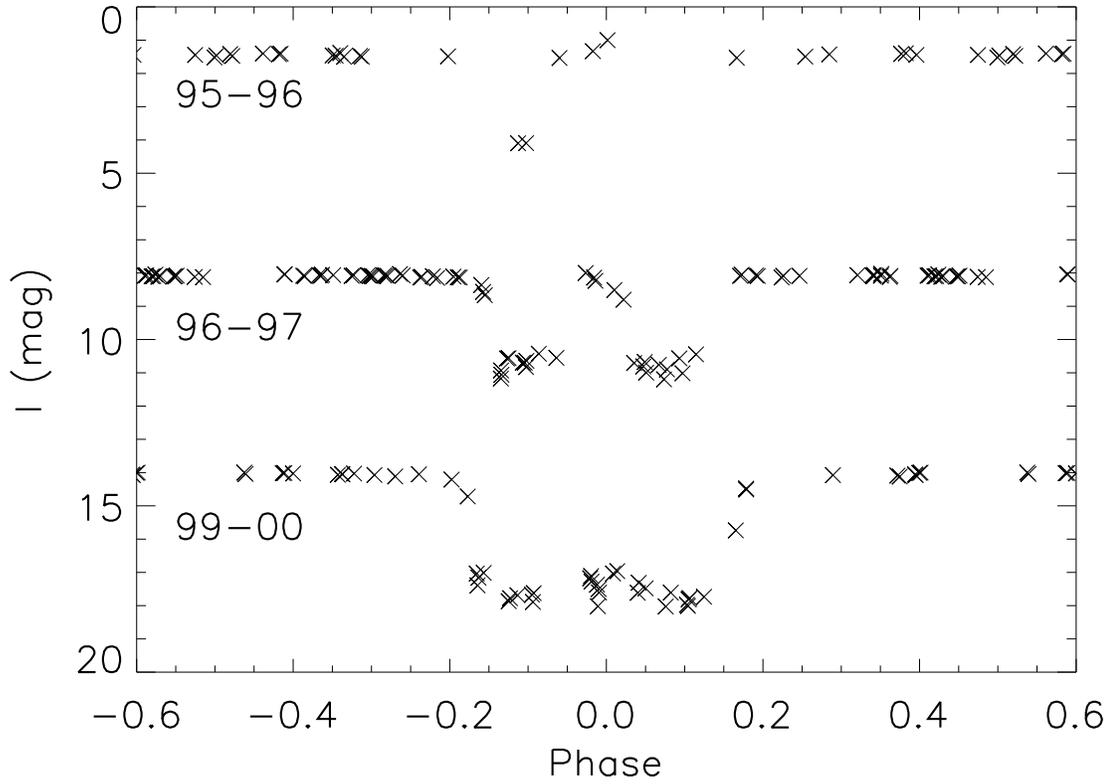}
\figcaption{ Light curves of KH15D during three different observing
seasons. All data were obtained with the 0.6 m telescope of Van Vleck
Observatory through a Cousins I filter. Differential magnitudes are
plotted with arbitrary offsets. It is clear that the width and depth
of the eclipse have increased with time and that the brightness of
the central peak has markedly decreased. \label{Fig. 2}}
\end{figure}

\begin{figure}
\plotone{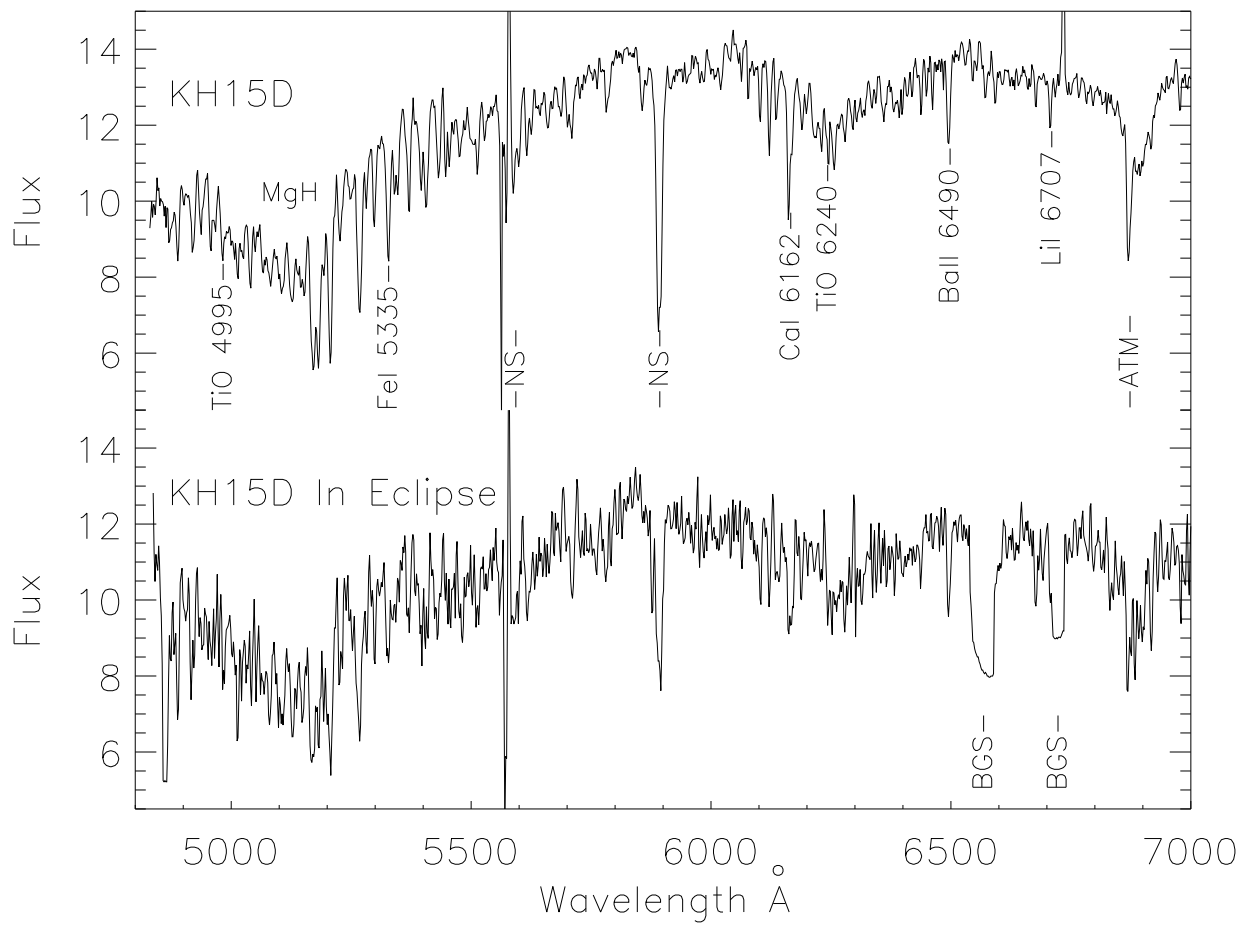}
\figcaption{A comparison of the spectrum of KH15D in and out of
eclipse.  The units of flux are 10$^{-16}$ erg cm$^{-2}$ s$^{-1}$.
The spectrum in the bottom panel was multiplied by a factor of 14.7 
in order to place it on the same scale as the spectrum obtained in the bright 
phase.  
Features used in the spectral classification of this object have been labeled.  
The truncated features marked BGS in the eclipse spectrum at 
H$\alpha$ and $\lambda$6717 [S II] are due to poor background subtraction 
and are not real features.  Also note the apparent emission of [S II] in 
the spectrum during its bright phase.  This is also an artifact of the 
background subtraction.  NS and ATM refer to the night sky and atmospheric
absorption features. \label{Fig. 3}}
\end{figure}

\end{document}